# THE EXPERIMENTAL STUDY OF THE SURFACE CURRENT EXCITATION BY A RELATIVISTIC ELECTRON ELECTROMAGNETIC FIELD [*]

G.A. NAUMENKO [†]

*Nuclear Physics Institute at Tomsk Polytechnic University, Lenina ave. 2a*
*Tomsky, 634050, Russua*

A.P. POTYLITSYN

*Tomsk Polytechnic University, Lenina ave. 2*
*Tomsky, 634050, Russua*

L.G. SUKHIKH

*Tomsk Polytechnic University, Lenina ave. 2*
*Tomsky, 634050, Russua*

YU. POPOV

*Tomsk Polytechnic University, Lenina ave. 2*
*Tomsky, 634050, Russua*

The problem of the surface current excitation in a conductive targets by a relativistic electron electric field as the origin of such radiation mechanisms as diffraction and transition radiation of relativistic electron was considered in frame of both surface current and pseudo-photon methods. The contradiction between these viewpoints in respect to the surface current on the target downstream surface necessitated the experimental test of this phenomenon. The test performed on electron beam of the 6 MeV microtron showed, that not any surface current is induced on the target downstream surface under the influence of a relativistic electron electromagnetic field in contrast to the upstream surface. This is important implication for the understanding of the forward transition and diffraction radiation nature.

## 1. About the problem

The association with a surface current is of weight conscious in many theoretical investigations of the forward diffraction radiation (FDR) and forward transition radiation (FTR) of relativistic electrons. Let's remind that diffraction radiation is the radiation emitted from the target-electron system when a relativistic electron moves rectilinary near the target close to the target edge. The transition and diffraction radiation (TR and DR) are of the same nature, but in the case of TR the electron crosses directly a target surface. We shall consider in this article only conductive targets, because for this case the properties of above-mentioned radiation types are more pronounced. There are at least two points of view on the radiation process for TR and DR: surface current approach and pseudo-photons approach.

---

[*] This work is supported by etc, etc.
[†] naumenko@npi.tpu.ru

### 1.1. *Surface current radiation viewpoint*

In this approach TR and DR are considered as a radiation of a current, induced on a conductive target surface by the relativistic particle electromagnetic field. This technique for DR from inclined conducting strip is described in [10]. In [9] the similar approach was used both for backward and forward TR and is irrespective to the thickness of the target surface.

For diffraction radiation this viewpoint was more pronounced in the article of B.M. Bolotovskiy [2], where he places primary emphasis upon the radiation formation length effect, when a charged particle moves over the conductive screen parallel to the axis $z$ with a velocity $v$ (Fig.1). It seems the formation length notion was introduced by M. Ter-Mikaelian in [1] as a result of interference between a charged particle electromagnetic field and a radiation field, emitted by this particle.

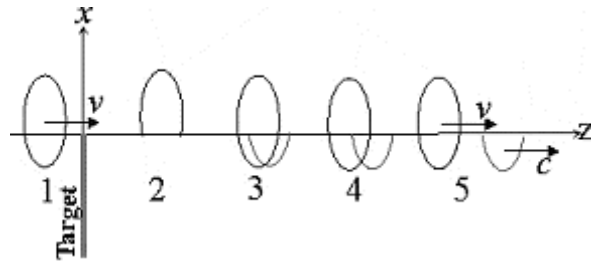

Figure 1. Illustration of the radiation formation (Fig. 6 from [2]).

According to [2] (as literally as possible) "position 1 in Fig. 1 shows the field of an initial charged particle. This field passes through the semi-surface and induces surface currents, which become sources of the DR. As the semi-surface is assumed to be ideally conductive one, the well known boundary conditions on the surface must be satisfied, namely, the total tangential electric field component on the target surface must be equal to zero. These conditions results the peculiar feature of the radiation field, namely, the radiation field is such one, that being close to the screen it neutralizes the part of a particle field, which attacks the screen."

Due to different velocity of a charged particle and FDR the particle field and FDR field separate then at the distance $\sim\gamma^2\lambda$, where $\gamma$ is the Lorentz-factor and $\lambda$ is the investigated wavelength. It is important, that this point of view assumes the obligatorily surface electric current on the target downstream surface irrespectively to the thickness of the conductive target.

The association between FDR and surface current is evident from the exact solution of Maxwell equations for diffraction radiation field from ideally conductive infinite thin semi-surface in [3], where the diffraction radiation field is expressed directly as a function of surface electric current:

$$A(R) = \frac{2\pi e^{i\omega R}}{R} J(k_0, q_0), \qquad (1)$$

where $k_0 = -\omega\sin\psi\cos\varphi$, $q_0 = -\omega\cos\psi$, $\varphi$ and $\psi$ are the angles of observation direction in spherical system and $J$ is the surface electric current density. However, this is not obvious for a thick target, if the target thickness is much larger than a skin-layer.

### 1.2. *Pseudo-photons reflection viewpoint*

On the other hand, the "pseudo-photon" method suggested by Fermi [4] and extended by Williams [5] is of considerable current use as the approach for electromagnetic processes theoretical investigations (see for instance [1] and [11]). According to this approach the charged particle field may be replaced by the field of photons, which in this case are named pseudo-photons (In [11] is used the term "virtual quanta". One should differ this term on one in quantum theory.). This approach provides a good accuracy for ultra-relativistic particles when the particle velocity becomes close to the light velocity ($v\rightarrow c$) (see [1]). In this case the longitudinal component of the particle electric field is negligible and the particle electromagnetic field being a transversal one has the same properties as a real photon field.

Before using of these properties first let's consider the interactions of real photons with the thick conductive mirror of a high reflectivity (the thickness is much larger than the skin-layer and the reflectivity is close to unit) in geometry, shown in Fig. 1. It is clear that the real photons are reflected almost fully, they don't penetrate through the target and they don't induce a surface current on the target downstream surface.

If the particle electromagnetic field and the real photons have the same properties, we may expect the same character of interactions of the particle electromagnetic field with a thick conductive mirror of a high reflectivity, namely, we may expect in contrast to the surface current viewpoint the absence of the surface electric current on the target downstream surface.

### 1.3. *Deduction*

Both these above-mentioned points of view are in principal contradiction in respect to the surface current induction and mechanism of FTR and FDR. The resolution of this contradiction may clarify the understanding of the nature of the FTR, FDR and other important phenomena. One of the ways of this problem resolution is an experimental test of the surface electric current on the downstream target surface.

There is not necessary to make an absolute surface current measurement, because we are sure, that backward transition and backward diffraction radiation (BDR) is emitted by the surface electric current on the upstream target surface. We may to make the relative measurements and compare a surface current on the downstream target surface with a surface current on the upstream one.

To exclude problems, which may be brought up due to the direct contact of relativistic electrons with a target material, the experiment in diffraction radiation (DR) geometry is preferable.

## 2. Experiment

### 2.1. *Experimental setup*

The experiment was carried out on the extracted electron beam of Tomsk Nuclear Physics Institute microtron with parameters, presented in Table 1.

Beam parameters of microtron allow to use coherent properties of a radiation to investigate in millimeter wavelength region DR characteristics and surface current induction. This possibility increases the surface electric current and diffraction radiation intensity by the 8 orders (proportional to the bunch population) and makes these values achievable for measurement using existing sensors.

Table 1. Microtron electron beam parameters

| Maximum energy | 6.2 MeV ($\gamma$ = 12.1) |
|---|---|
| Bunch length | $\sigma_l \approx$ 1.4 mm |
| Single-bunch population | N=$10^8$ |
| Number of bunches in macro pulse | n=$10^4$ |
| Macro pulse duration | 4 $\mu$sec |
| Macro pulse repetition | 3-10 Hz |
| Beam size in extraction point | $\sigma_l \approx$ 2.3 mm |
| Beam divergence | $\sigma_d \approx$ 0.046 radians |

For the surface electric current measurement we use the well-known technique, which is applied for the surface current measurement in strip-line beam position monitors [6]. Fig. 2 shows the surface electric current sensor built-in diffraction radiation target.

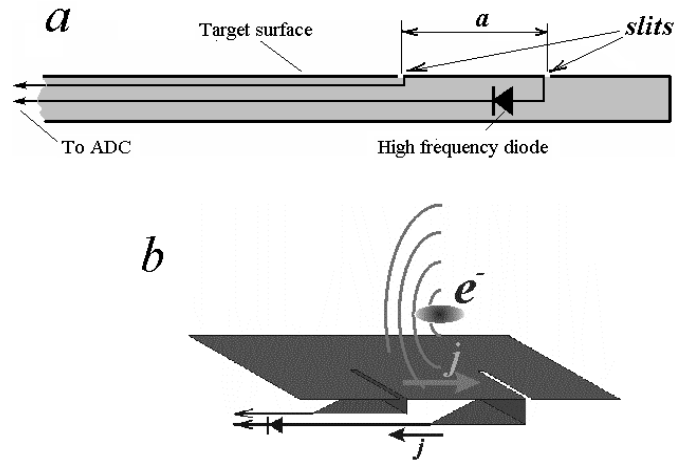

Figure 2. Surface electric current sensor built-in diffraction radiation target. *a* – the target part including the surface electric current sensor. *b* – surface current sensor cross section.

To have a similarity of BDR and surface current on impact-parameter the sensor strip length *a* was chosen 3 mm (the quarter of the average wavelength of registered BDR). The sensitivity of sensor is ≈50 mV/[mA/cm] (estimated lower limit). The slit width was 0.4 mm.

The DR target with built-in surface current sensor was tested on the real photon beam from the pulse emitter with 6.7 mm wavelength radiation. On Fig's 3 and 4 are shown the scheme of experiment and layout of experimental equipment, used for this test.

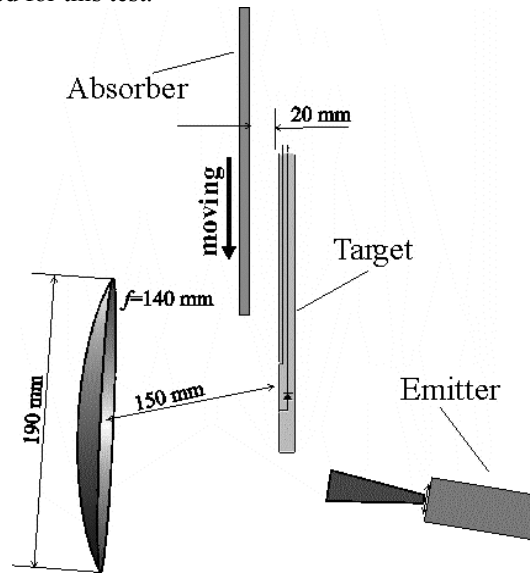

Figure 3. The acheme of experiment for test of the surface electric current sensor built-in diffraction radiation target on the real photon beam.

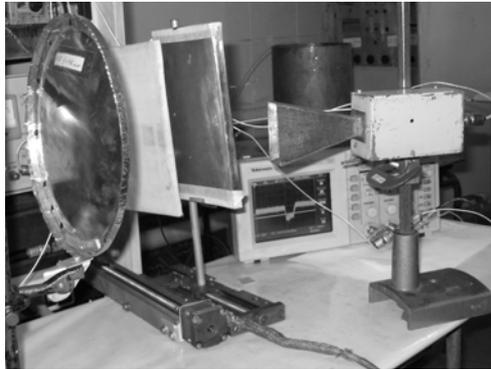

Figure 4. Layout of experimental setup for test of surface electric current sensor built-in diffraction radiation target on the real photon beam.

The radiation from emitter was focused on the sensor by parabolic mirror. The movable absorber allows to screen the sensor step by step.

Fig. 5 shows the dependencies of the measured surface current on the absorber position for the upstream and downstream position of the sensor on the target. We can see, that in case of real photons a surface electric current on the downstream target surface actually absent.

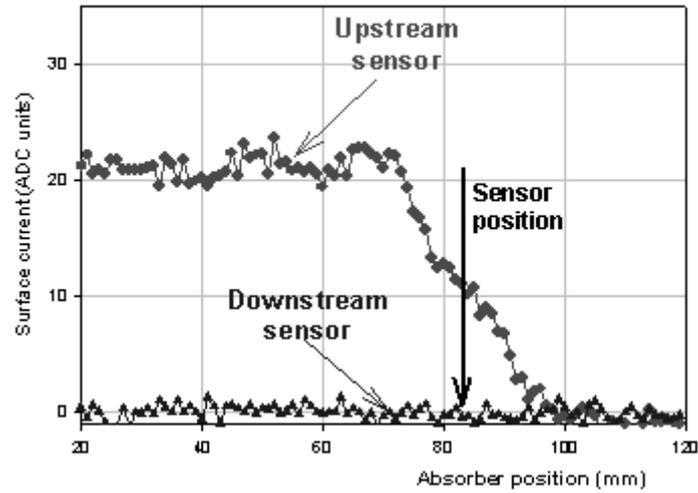

Figure 5. Dependence of the surface current sensor response on the absorber position in case of upstream and downstream sensor position, measured on the real photon beam.

Since BDR is generated by the electric current on the upstream target surface, we may check the correlation between the surface current and BDR intensity by simultaneous measurement the surface current and BDR. Therefore the scheme of experiment (Fig. 6) was provided the measurement of BDR and surface current simultaneously.

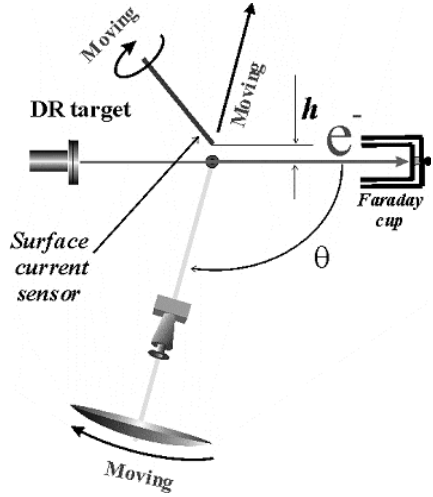

Figure 6. Scheme of experiment for study of the surface current induction under the influence of a relativistic electron electromagnetic field.

For the BDR measurement we had used the method based on observation using the parabolic telescope, described in [7], which provides the measurement a BDR in far field zone mode. The parabolic telescope may be rotated around the vertical axes for measurement of the horizontal BDR angular distribution (Fig. 7*b*) and for tuning the horizontal angle of the BDR for center of the horizontal distribution.

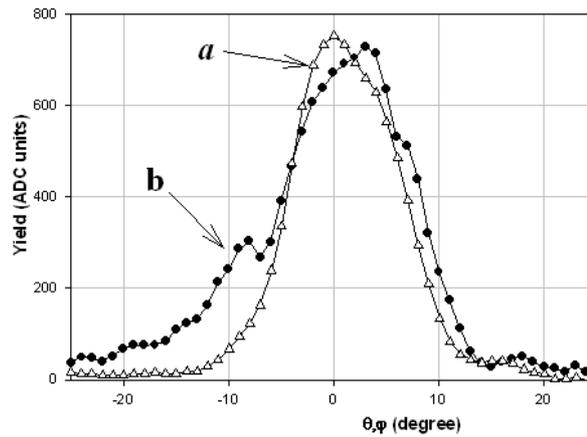

Figure 7. Vertical (*a*) and horizontal (*b*) angular distribution of BDR.

The target with the built-in surface electric current sensor may be moved in horizontal plane for changing the impact-parameter $h$ and may be rotated around the horizontal axes for measurement of the vertical BDR angular distribution (Fig. 7*a*) and for tuning the vertical angle of the BDR for center of this distribution. For the background measurement from surface current sensor we glued up the strip of sensor by the conductive foil using conductive glue.

The Faraday cup signal allows to measure an impact parameter using the electron beam intensity suppression when electrons hit the target.

## 2.2. *Results*

The measurements were done for upstream and downstream orientations of surface current sensor and results were compared for both cases in the same units. The measured simultaneously a surface current and BDR intensity as a function of impact-parameter for both orientations of sensor (upstream and downstream) after background subtraction are presented In Fig. 8. Maximal surface electric current density was 0.15 mA/cm (estimated lower limit).

.

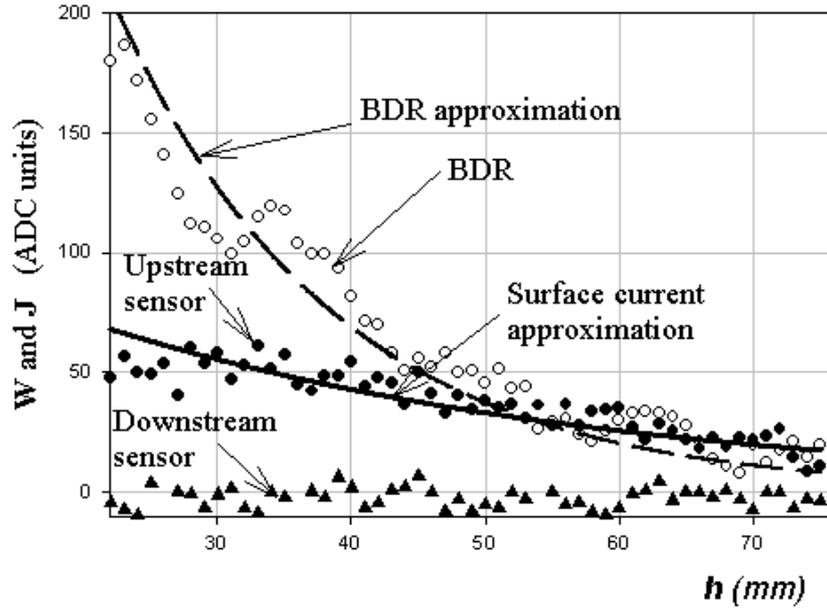

Figure 8. Measured dependence on impact-parameter of the BDR and surface electric current for upstream and downstream orientation of the current sensor.

## 3. Summary

In the considered geometry BDR intensity has an exponential dependence on the impact-parameter $\sim A \cdot e^{-4\pi h/\gamma\lambda}$ [8]. In Fig. 8 the dashed line presents the approximation of the experimental BDR dependence by this exponent with next parameters: $\gamma$=12.1, A=808±33, $\lambda$=17±3mm. The oscillations in experimental dependence are caused by finite size of the target width. In case of upstream sensor orientation we see in Fig. 8 the correlation of the dependences on the impact-parameter between a surface electric current and BDR, which confirms the induction of the surface current by an electron bunch field. According to [3] a surface current has also an exponential dependence on the impact-parameter $\sim B \cdot e^{-2\pi h/\gamma\lambda}$. In Fig. 8 the solid line presents the approximation of the experimental surface current dependence with next parameters: $\gamma$=12.1, B=120±21, $\lambda$=20±4mm. However for the downstream sensor orientation the value of surface current is equal zero within the limits of experimental error. So we can claim than with accuracy to the experimental error the surface current on the downstream surface is absent. This is an important result of this experiment, because it defines an understanding of a nature of the forward diffraction and transition radiation from conductive targets.

**Acknowledgments**

**References**

1. M.L. Ter-Mikaelian, High Energy Electromagnetic Processes in Condensed Media, Wiley-Interscience, New York, 1972.
2. B. M. BolotovskiI. Preprints of Lebedev Institute of Physics, Russian Academy of Sciences,Vol 140 (1982), p. 95
3. A.P. Kazantsev, G.I. Surdutovich, Dokl. Akad. Nauk SSSR 147 (1962).
4. E. Fermi, Z. Phys., 29, 315 (1924)
5. E. Wiliams, K. Danske Vidensk. Selsk., 13, 4 (1935)
6. V. Sargsyan, "Comparison of Stripline and Cavity Beam Position Monitors", TESLA Report 2004-03.
7. B. N. Kalinin, G. A. Naumenko, A. P. Potylitsyn,JETP Letters, 2006, Vol. 84, No. 3, pp. 110–114.
8. A. Potyitsyn, NIM B 145 (1998) 169
9. S. Reiche, J. B. Rosenzweig, Transition Radiation for Uneven, Limited Surfaces. Proceedings of the 2001 Particle Accelerator Conference, Chicago.
10. J.H. Brownell and J. Walsh. Phys. Rev. E 57,1 (1998) 1075
11. J.D. Jackson, Classical Electrodynamics, 3$^{rd}$ ed. J. Willey&Sons, New-York, 1998.